\documentclass[conference]{IEEEtran}
\IEEEoverridecommandlockouts

\usepackage{cite}
\usepackage{amsmath,amssymb,amsfonts}
\usepackage{algorithmic}
\usepackage{graphicx}
\usepackage{textcomp}
\usepackage{xcolor}
\usepackage{epstopdf}
\usepackage{amsthm}
\usepackage{nicefrac}
\usepackage{balance}
\usepackage{tabularx,booktabs}
\usepackage{color}
\usepackage{tikz}
\usetikzlibrary{arrows.meta}
\usepackage{subcaption}
\usepackage{mwe}

\theoremstyle{definition}

\usepackage[nodisplayskipstretch]{setspace}
\begin{document}

\title{Low-Latency Communication with Computational Complexity Constraints\\
\thanks{This work was funded in part by the Swedish foundation for strategic
research.}
}

\author{
\IEEEauthorblockN{Hasan Basri Celebi, Antonios Pitarokoilis, Mikael Skoglund}
\IEEEauthorblockA{
School of Electrical Engineering and Computer Science
\\
KTH Royal Institute of Technology, Stockholm, Sweden
}
}

\maketitle

%%%%%%%%%%%%%%%%%%%%%%%%%%%%%%%%%%%%%%%%%%%%%%%%%%%%%%%%%%%%%%%%%%%%%%%%%%%%%
\begin{abstract}
Low-latency communication is one of the most important application scenarios in next-generation wireless networks. Often in communication-theoretic studies latency is defined as the time required for the transmission of a packet over a channel. However, with very stringent latency requirements and complexity constrained receivers, the time required for the decoding of the packet cannot be ignored and must be included in the total latency analysis through accurate modeling. In this paper, we first present a way to calculate decoding time using \textit{per bit} complexity metric and introduce an empirical model that accurately describes the trade-off between the decoding complexity versus the performance of state-of-the-art codes. By considering various communication parameters, we show that including the decoding time in latency analyses has a significant effect on the optimum selection of parameters. 
%Finally, based on the results presented, maximization of the total number of information bits to be transmitted under a stringent latency requirement is discussed.
\end{abstract}

\begin{IEEEkeywords}
Low-complexity receivers, low-latency, IoT, channel coding, ordered statistics decoder
\end{IEEEkeywords}

\section{Introduction}

Ultra-reliable, low-latency communications (URLLC) have recently attracted significant research interest, due to emerging delay-critical applications, such as machine-to-machine communication, remote medical surgery, factory automation, and automated traffic control \cite{ji_ultra_reliable}. The performance of latency-constrained communication systems has mostly been evaluated in terms of outage probability, i.e., the probability that the instantaneous mutual information falls below a desired rate. However, it has been shown that, for short packets such an approximation can provide inaccurate estimates since it becomes valid when the length of the transmitted packet grows very large. Accurate bounds on the maximal coding rate that are non-asymptotic with respect to the length of the transmitted packet were given in \cite{polyanskiy_channel_coding}.

Both the outage probability and the non-asymptotic bounds assume that decoding happens instantaneously, i.e., the time required for the decoding of a packet is negligible \cite{grover_fundamental_limits}. This assumption can be justified for small transmission rates, unlimited computational power at the decoder side or loose latency requirements \cite{rachinger_comparison_of}. However, for low-latency communication with complexity-constrained receivers, such as low-budget IoT devices, due to the slower processor capabilities, the decoding time must be accurately modeled and included in the total latency analysis.

%Physical layer latency of a point-to-point communication system can be divided into several parts, such as encoding, transmit, propagation, processing latencies, and finally the decoding latency. Accordingly, in order to decrease each of these latencies, various researches are done on packet structures, scheduling, block-length and coding rate selection, practical codes, etc. (see \cite{ji_ultra_reliable, shivarnimoghaddam_short_block, niu_polar_codes} and references therein). However, optimizing the performance of a system by improving some aspects always causes a cost on another one. Such trade-offs need to be carefully identified and deeply investigated since they may reveal the limitations of the system which are not trivial beforehand.

To the best of the authors' knowledge, there is no generally accepted model of the computational complexity of a typical channel decoder. It appears, however, the number of operations per information bit is often selected as a metric for the computational complexity \cite{khandekar_on_the}. A brief summary on decoding complexities of practical codes is presented in \cite{niu_polar_codes}. An overview on the recent developments for short block-length codes is made in \cite{liva_code_design} where the authors also discuss the decoding complexities of several state-of-the-art decoders. It is remarked that codes which approach the theoretical limits require higher computational complexities. Recently, in \cite{sybis_channel_coding} the authors study the computational complexity, defined as the total number of binary operations, of some practical decoders and compare their block-error-rate (BLER) performance for short block-lengths. In \cite{shivarnimoghaddam_short_block} the \emph{per information bit} computational complexity, i.e., number of binary operations per information bit, is studied as a function of BLER. It is further shown that an excess power with respect to the normal approximation in \cite{polyanskiy_channel_coding} must be spent to achieve a fixed allowed BLER at a fixed code-rate, when a particular code is chosen.

In this paper we propose a comprehensive model that relates various parameters of low-latency communication systems with computational complexity constraints, such as decoding complexity (in number of binary operations per information bit), BLER, signal-to-noise ratio (SNR), code rate, and codeword block-length, in an accurate and tractable way. In particular, we first consider the complexity of an ordered statistics (OS) decoder and derive a bound on the complexity that is mathematically tractable. A consistent way to calculate the decoding time of a complexity constrained receiver is presented. Then, a model which reveals the trade-off between the complexity versus performance, in terms of BLER, of OS decoders is proposed. Using the model, the minimum amount of power penalty that is required to meet the reliability constraint is derived. Finally, based on the proposed model we study some interesting communication scenarios that reveal the effect of decoding complexity in latency constrained communication.

\section{System Model}\label{sec_system_model}

Communication over a binary-input, additive white Gaussian noise (BI-AWGN) channel is considered. Let $x_i$ be the input symbol at the $i$-th channel use of duration $T_s$ seconds, selected from the set $\{-1,+1\}$. Then, the received sample $y_i\in\mathbb{R}$ is given by
\begin{align}\label{eq_system_model}
y_i = x_i + z_i,
\end{align}
where $z_i\sim\mathcal{N}(0,\sigma^2)$ and the signal-to-noise ratio is $\rho=\sigma^{-2}$. Transmission occurs in codewords of $n$ symbols and the transmission duration is $d_T=nT_s$ seconds. Each codeword is the output of the channel encoder at an input of $k$ information bits. Hence, the information rate of the code is $r=\frac{k}{n}\in(0,1]$.

In this work, the normal approximation to bounds from finite blocklength information theory is used as the benchmark for the maximum information rate over the BI-AWGN \cite{polyanskiy_channel_coding}. The normal approximation for the channel in \eqref{eq_system_model} for a codeword of length $n$, with a codeword error probability not exceeding $\epsilon\in(0,1)$ is given by
\begin{equation}\label{eq_normal_approximation}
R(n,\epsilon,\rho)=C(\rho)-\sqrt{\frac{V(\rho)}{n}}Q^{-1}(\epsilon)\log_2e+O\left(\frac{1}{n}\right),
\end{equation}
where $C(\rho)$ denotes the capacity, $V(\rho)$ the dispersion and $Q^{-1}(\cdot)$ is the inverse to the $Q$-function \cite{erseghe_coding_in}.

\section{Computational Complexity of Decoders}

Given a codeword of $n$ symbols, the total communication latency is given by
\begin{align}\label{eq_total_latency}
d_t=nT_s + d_D,
\end{align}
where $d_D$ is the time required for the decoding. The goal of this section is to provide a mathematical expression for $d_D$, that can summarize in a simple and intuitive way various parameters of decoding algorithms that influence the decoding time. 

\subsection{Maximum Likelihood Decoder} 

The maximum likelihood decoder, which minimizes the probability of codeword error for equiprobable codewords, compares the vector of observations with every codeword and decides for the one minimizing the Euclidean distance. Since there are $M=2^k$ codewords, the number of operations (not necessarily binary) per information bit is $M/k$, which implies that the complexity of the optimal decoder is exponential in $k$. This can be considered as an upper bound on the computational complexity of any practical code.

A simple expression that can exactly characterize the computational complexity of every code is unlikely to be found. In this work, we propose a model for the computational complexity that is based on using Bose, Chaudhuri and Hocquenghem (BCH) codes with OS decoders. Before justifying this choice a brief description of the OS decoders is in order.

\subsection{Ordered Statistics (OS) Decoding} 

OS decoding is a near-ML soft-decision universal decoding algorithm for any $(n,k,d_{\min})$ linear block code, where $d_{\min}$ is the minimum Hamming distance of the code. For a given observation vector $\boldsymbol{y}$, the log-likelihood ratios (LLRs) are computed and sorted in descending order. The generator matrix is permuted in the corresponding way and transformed to systematic form via Gauss--Jordan elimination. A detailed exposition of the OS decoding can be found in \cite{fossorier_soft_decision}. Here, we will focus on the complexity. 

The key parameter of OS decoders is the order, $s \in \mathbb{R}^+$, that is intimately related with the computational complexity of the algorithm. Let $E_s$ be the set of error patterns with Hamming weight up to $s$. These are the error patterns that are checked by the OS decoding algorithm. When $s$ is small only a few error patterns are checked and the error rate is high. As $s$ increases, the computational complexity increases and the performance approaches the ML decoder. The cardinality of $E_s$ is $ \left|E_s\right| = \sum_{i=0}^s  {{k}\choose{i}} $. In fact, this number is the total number of codeword comparisons and can be very high even for low $s$. For the AWGN channel, the recommended $s$ for near-ML performance is \cite{fossorier_soft_decision}
\begin{equation}\label{eq_recommended_complexity_order}
s_r = \min\left\lbrace\left\lceil \frac{d_{\min}}{4} - 1 \right\rceil, k  \right\rbrace, 
\end{equation}
where $\lceil\cdot\rceil$ is the ceiling function.

\subsection{Computational Complexity}

The computational complexity of BCH codes with OS decoding is a reasonable choice for the modeling of the computational complexity of more general codes for various reasons. In \cite{shivarnimoghaddam_short_block} it was shown that the extended BCH (eBCH) codes with OS decoders come very close to the normal approximation \eqref{eq_normal_approximation} for short block-length codes. The OS decoding algorithm allows for a simple parameterization of the decoding complexity via a single parameter, i.e., the order $s$. In \cite{shivarnimoghaddam_short_block} it was also shown that as $s$ increases, the computational complexity rapidly increases and the performance approaches the ML. On the other hand, for small $s$ the computational complexity is reduced and the performance gracefully degrades. Further, the BCH codes are reasonably flexible in terms of the choice of the coding rate.

Focusing on the computation-intensive operations performed by the OS decoding algorithm can provide an immediate estimate of the total number of binary operations per information bit, i.e.,
\begin{equation}
c =  \frac{k^2}{8} +
\frac{n}{2} \sum_{i=0}^s  {{k}\choose{i}} , 
\label{eq_exact_number_of_bin_operations}
\end{equation}
where the first term is due to the Gauss-Jordan elimination of the permuted generator matrix and the second term is the sum of the vector multiplications and comparisons for each error pattern \cite{shivarnimoghaddam_short_block}. Note that, for $s\leq 2$ the computational complexity is dominated by the first term, otherwise the second term dominates the complexity. Thus, the order of the complexity can be shown as
\begin{equation}
	c = \begin{cases}
	\mathcal{O}(k^2), & \text{if $s \leq 2$},\\
	\mathcal{O}(k^s), & \text{if $s > 2$}.
	\end{cases}  
\end{equation}

It is worth to note that (\ref{eq_exact_number_of_bin_operations}) is hard to interpret and increasingly difficult for mathematical tractability due to the sum of binomial coefficients. Therefore, we derive the following upper bound on $ c $ for further analyses.
\begin{equation} 
c \leq \frac{k^2}{8} + \frac{n}{2} 2^{k h\left(\frac{s}{k}\right)}
\label{eq_complexity_bound}
\end{equation}
where $h(z) = -z\log_2(z)-(1-z)\log_2(1-z)$ is the binary entropy function. See Appendix for the proof. Note that (\ref{eq_complexity_bound}) gets tighter with higher values of $ s $.

%Finally, it is debatable whether the chosen complexity measure is the most reasonable choice.  However, since we focus on decoding time, the algorithmic complexity, which can be represented best with the number of computational operations, is chosen as the complexity metric.

\subsection{Decoding Duration}

Let $T_b$ be the time required for a binary operation on the hardware platform that the decoder operates. Then the total latency for the transmission of a codeword of blocklength $n$ is given by
\begin{equation}
d_t = n T_s + d_D = n T_s + k c T_b.
\label{eq_total_latency_with_c}
\end{equation}

The decoding time is influenced by the particular hardware platform. For simplicity and generality we assume a linear relation between $ d_D $ and $ T_b $. Accuracy of $d_D$ can be improved further by evaluating the hardware technology. But since this is beyond the scope of this paper, we confine to (\ref{eq_total_latency_with_c}) for further analysis. 

Suppose that a latency constraint on $ d_t $ is applied such as
\begin{equation}
d_t =	n T_s + k c T_b \leq d_m,
\label{eq_total_latency}
\end{equation}
where $d_m$ is the maximum latency deadline for $ d_t $. Such a latency constraint restricts $ s $ as follows
\begin{equation}
	s_m = \underset{\{s| s \in \mathbb{Q^+}, ~nT_s+ckT_b\leq d_m\}}{\mathrm{arg~max}} c,
\end{equation}
where $ s_m $ denotes the maximum allowed order. Hence, using (\ref{eq_complexity_bound}) and (\ref{eq_total_latency}), the following inequality follows
\begin{equation}\label{eq_bound_on_order}
h\left(\frac{s}{k}\right) \leq \frac{1}{k}\log_2 \gamma,
\end{equation}
where $ \gamma = \frac{8d_m - 8nT_s - k^3T_b}{4nkT_b} $. Note that (\ref{eq_bound_on_order}) is not an upper bound on $ s $, but meeting (\ref{eq_bound_on_order}) guarantees the latency deadline. By using a tight approximation for binary entropy function, given as $ h(q) \approx (4 q (1-q))^{3/4} $, $ s_m $ can be approximated as
\begin{equation}
s_m \approx \frac{k}{2} \left( 1 - \sqrt{1 - \left( \frac{\log_2 \gamma}{k} \right)^{4/3}} \right).
\label{eq_bound_on_t}
\end{equation}

\section{Computational Complexity vs Power Penalty}

The selection of an order $s$ for a particular code of fixed $n$, $k$ and $\rho$ can be used to control the total latency $d_t$ of the communication, albeit at the expense of reduced reliability. One way to satisfy a desired target of reliability of $\epsilon$, i.e., codeword error probability, a power penalty has to be paid. Hence, an interesting, yet complex, relation between total latency, computational complexity of the decoder and power spent for transmission arises. 

\begin{figure}[t]
	\centering
	\includegraphics[trim=0 0 0 0, scale=0.62]{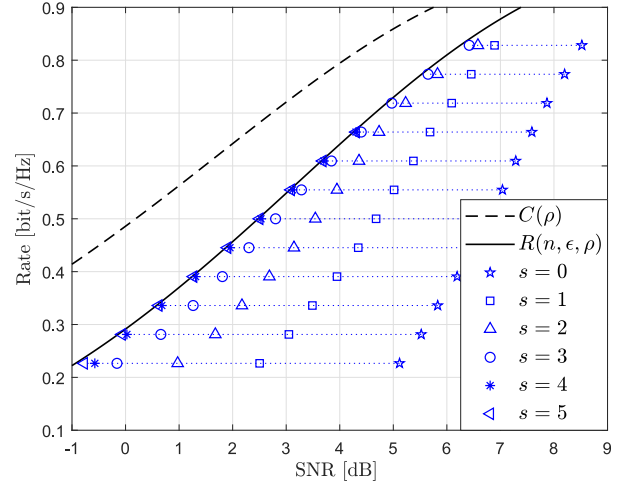}
	\caption{Power requirements of OS decoders with different orders at different rates for $\epsilon=10^{-3}$ when $n=128$.}
	\label{fig_power_vs_rate_w_decoders}
\end{figure}

In general, a direct proportion between complexity and BLER performance is expected for a decoder. Thus, a decoder will perform better when its complexity increases, and vice versa. Empirical results shown in \cite{shivarnimoghaddam_short_block} reveal that the complexity of the OS decoder exponentially increases as its performance approaches to the normal approximation. Therefore, in this section, we aim to model the trade-off between computational complexity and performance of OS decoders in the finite block-length regime with a tractable expression. Consequently, we first analyze the performance of OS decoders over BI-AWGN channel with different orders at different coding rates for $ n=\{64,128\} $ where the information bits are encoded with eBCH encoder. Results for $ n=128 $ are plotted in Fig. \ref{fig_power_vs_rate_w_decoders} where the rate as a function of the SNR in dB is plotted. The dashed line is the ergodic capacity in the asymptotic regime, the solid line is the normal approximation \eqref{eq_normal_approximation} for $n=128$ and $\epsilon=10^{-3}$. Each blue, horizontal line joins the operating points of eBCH codes with OS decoding, BLER $\epsilon=10^{-3}$ and a fixed rate. Starting from right to left the order of the decoder increases from $s=0$ to $s=5$ and each marker shows the required SNR of a decoder at a fixed rate. Note that these SNR values are computed by starting the BLER analyses from low SNR and detecting the required amount by gradually increasing it until BLER reaches $ \epsilon $. Fig. \ref{fig_power_vs_rate_w_decoders} illustrates some significant results. First of all, it shows that performance of OS decoders achieve $ R(n,\epsilon,\rho) $ at any rate if $ s $ is sufficiently high. It is also clear that as the computational complexity of the decoder increases with increasing $s$, the power penalty required for the desired codeword error probability decreases. Conversely, a computational constraint due to stringent total latency constraints leads to a corresponding power penalty. Similar results are obtained from the analyses when $ n=64 $. However, these results are not shown due to space-limitations.

\begin{figure}[t]
	\centering
	\includegraphics[trim=0 0 0 0, scale=0.62]{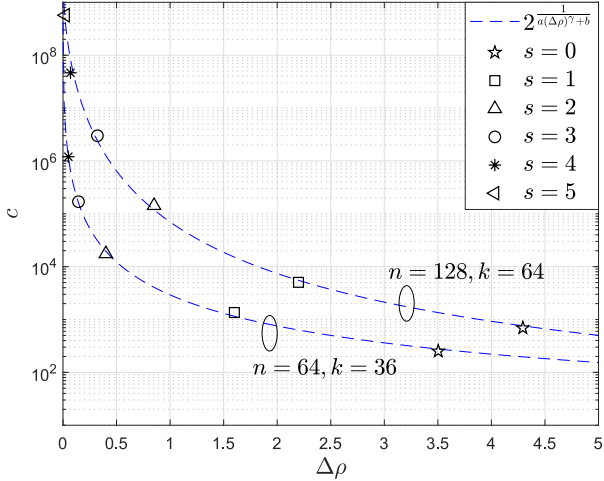}
	\caption{Comparison of the proposed model with actual results.}
	\label{fig_q_vs_deltaP_k_64}
\end{figure}

In Fig. \ref{fig_q_vs_deltaP_k_64} the number of binary operations per bit, $c$, is plotted as a function of the power penalty $\Delta\rho$ for two codes with blocklength $n=\{64, 128\}$ and $k=\{36, 64\}$, respectively. The individual points correspond to simulation results with decoders of order $s=\{0,1,\ldots,5\}$. It can be observed that in both cases, the relation between computational complexity and power penalty is closely approximated by a law of the type
\begin{equation}
	\log_2 c = \frac{1}{a ({\Delta}\rho)^\gamma + b},
	\label{eq_q_delta_tradeoff}
\end{equation}
for appropriate choices of the positive constants $\alpha$, $\gamma$ and $b$. This law is plotted as the dashed lines. For the case of $n=64$ the parameters were found to be $ a=0.05 $, $ b=0.03 $, $ \gamma=0.4 $ and for $n=128$ the parameters are given by $ a=0.03 $, $ b=0.03 $, $ \gamma=0.6 $.

The expression in \eqref{eq_q_delta_tradeoff} summarizes in a simple way an intuitive trade-off between computational complexity and power penalty for a fixed reliability constraint. Even though it was derived based on the OS decoder, numerical results in previous works \cite[Fig. 6]{shivarnimoghaddam_short_block} show that other families of codes, such as tail-biting convolutional codes and polar codes, follow a similar law when it comes to the relation between computational complexity and power penalty. Hence, it can be advocated that \eqref{eq_q_delta_tradeoff} is a useful proxy for the study of URLLC systems with computational complexity constraints.

\section{Numerical Examples}

In this section we present three interesting communication scenarios with latency, reliability and computational complexity constraints. These scenarios reveal that when the decoding time is included in the latency modeling of URLLC communications, the choice of optimal operating point becomes a non-trivial task and gives rise to an abundance of interesting problem formulations.

\subsection{Maximal Information Rate}

\begin{figure}[t]
	\centering
	\includegraphics[trim=0 0 0 0, scale=0.62] {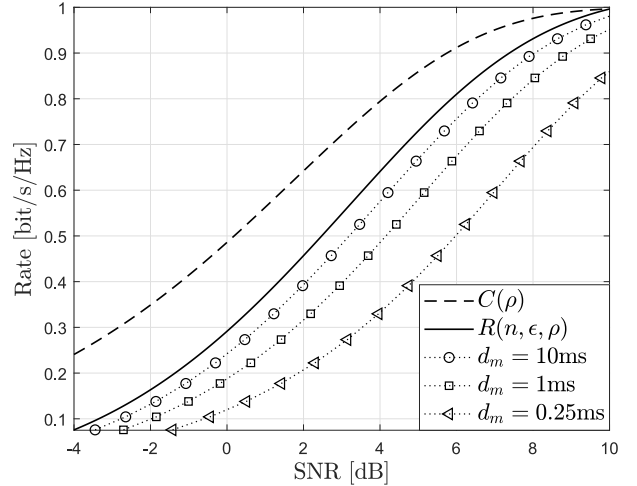}
	\caption{New achievability bounds  under latency and complexity constraints for $ n=128 $, $ \epsilon = 10^{-3} $, $ T_s = 1\,\mu$s, and $ T_b=1\,$ns.}
	\label{fig_power_vs_rate_w_timeconst}
\end{figure}

In Fig. \ref{fig_power_vs_rate_w_timeconst} the information rate is plotted as a function of the SNR in dB. The dashed line corresponds to the ergodic capacity and the solid line to the normal approximation \eqref{eq_normal_approximation} for $n=128$ and $\epsilon=10^{-3}$. The remaining three plots correspond to maximal information rate when a total latency constraint of $d_m=\{10,1,0.25\}\,$ms is imposed. It is assumed that the symbol interval is $T_s=1\,\mu$s and the time required for a binary operation is $T_b=1\,$ns. In particular, for each rate and blocklength, $n$, the maximum allowable decoding time is calculated using \eqref{eq_total_latency_with_c}. This in turn yields the required power penalty $\Delta\rho$ via \eqref{eq_q_delta_tradeoff}. Finally, the point on the plot is determined by shifting the point on the normal approximation by $\Delta\rho$ to the right.

\subsection{Maximization of $ k $}

We consider the case that a message is subject to a total latency constraint of $d_m$, with codeword error probability $\epsilon$ and there is also a total power budget of $P_m$ at the transmitter. We intend to maximize the number of information bits $k$ that can be transmitted within the total latency and power budget. 

When there is unlimited computational power, a codeword can be decoded instantaneously and therefore all the total latency budget can be used for the transmission of the message, i.e., $n=d_m/T_s$ symbols can be transmitted at a rate that is determined by \eqref{eq_normal_approximation}, $R(d_m/T_s,\epsilon,P_m)$, which yields
\begin{equation}
	k_m = \left\lfloor \frac{d_mR(d_m/T_s,\epsilon,P_m)}{T_s} \right\rfloor .
\end{equation}

However, for a computational complexity constrained receiver an interesting trade-off arises. If $n$ is selected small, the available duration for decoding can be sufficient so that a high rate code can be used. As $n$ increases, the available duration for decoding shrinks and a code with decreasing coderate must be selected so that the total latency constraint is satisfied.

\begin{figure}[t]
	\centering
	\includegraphics[trim=0 0 0 0, scale=0.62]{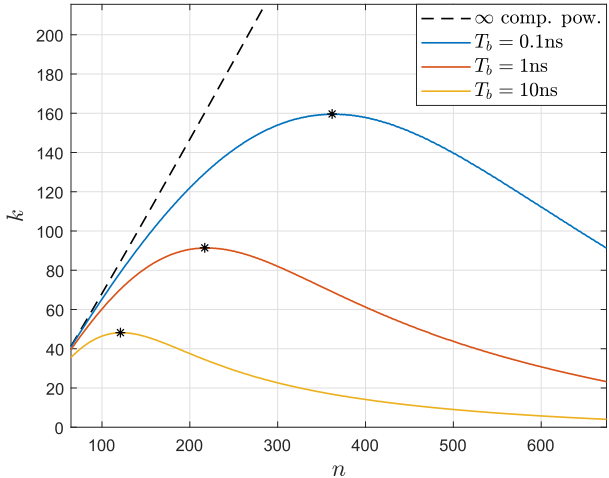}
	\caption{Maximum $ k $ for several complexity constrained receivers where $d_m = 1\,$ms, $ P_m=5\, $dB, and $\epsilon=10^{-3}$.}
	\label{fig_k_maximization}
\end{figure}

In Fig. \ref{fig_k_maximization} numerical results that correspond to the investigated scenario are plotted for  $ d_m = 1\, $ms, $ P_m=5\,$dB, and $ \epsilon=10^{-3} $. Three different choices for execution times for a binary operation are shown that correspond to $ T_b = \{10, 1, 0.1, 0 \} $ns, where $ T_b = 0\, $ns stands for infinite computation power. The previously introduced trade-off is clear here and the maximums appear at $ n = \{ 121, 217, 362, 1000\} $, respectively. Corresponding $ k_m $ values are $ k_m = \{48, 91, 159, 803\}$. 

Note that ratios of $ k_m $ values found for complexity constrained receivers to the $ k_m $ of infinite computation power receiver are $ 0.06, 0.11, 0.2 $, respectively. Thus, one can conclude that if complexity constraints and decoding duration are taken into account, one can transmit even less than 20\% of the theoretical limits, depending on the receiver capabilities.

\subsection{Minimization of $ d_t $}

\begin{figure}[t]
	\centering
	\includegraphics[trim=0 0 0 0, scale=0.62]{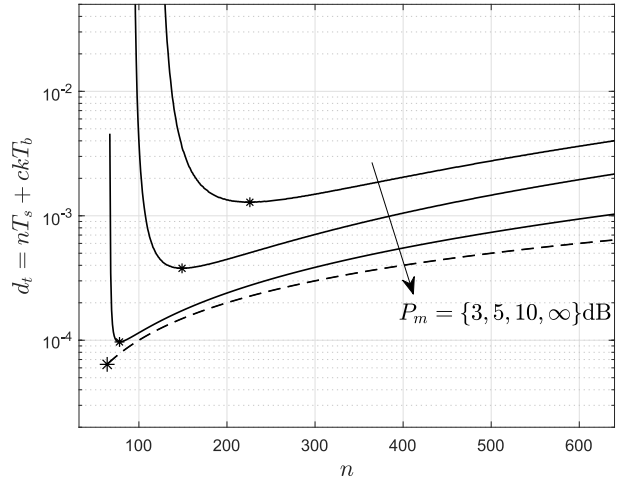}
	\caption{Minimum $ d_t $ with respect to $ n $ for several $ P_m $ where $ k=64 $, $\epsilon=10^{-3}$, and $ T_b = 10^{-9}\, $s.}
	\label{fig_t_t_minimization}
\end{figure}

Another interesting trade-off arises when we intend to transmit a fixed number of information symbols, $k$, subject to a codeword error probability constraint, $\epsilon$ and a maximum power constraint, $P_m$. In Fig. \ref{fig_t_t_minimization} the total latency is plotted as a function of the codeword length, $n$. It can be seen that for small $n$ the coderate of the selected codebook must be very high. Hence, either the transmission is not possible when the required coderate exceeds \eqref{eq_normal_approximation} or the required decoder must operate very close to the normal approximation, which yields a very high required computational complexity. This translates to very high total latency. As $n$ increases, the required rate is decreasing, hence it is more likely that it can be supported by the power budget or a rate sufficiently far from the normal approximation can be selected. In this case, a decoder with low complexity can be selected and the total latency is dominated by the codeword transmission latency. In Fig. \ref{fig_t_t_minimization} the optimal codeword lengths are $ n_{\text{opt}} = \{226, 149, 78\} $ for power constraints $ P_m = \{3, 5, 10\}\,$dB, respectively. Infinite $P_m$ implies that the symbols are transmitted error free and $ n_{\text{opt}} = k $ since from (\ref{eq_q_delta_tradeoff}), $ d_D = kT_b \approx 0 \,$s and hence $ d_t = nT_s $ and linearly increases in $ n $.

\section{Conclusion} 

In this study, we focus on latency caused by signal transmission and decoding. We investigate their effect on the transmission parameters by modeling the behavior of OS decoders with an accurate and tractable mathematical expression. Results show that decoding time has a considerable effect on the bounds of the short block-length codes if there is a latency constraint on the system. It is shown that if complexity constraints and decoding duration are considered in the maximization of the total number of information bits to be sent under latency and error rate constraints, less than 20\% can be achievable compared to the theoretical limits.

\appendix

	Let $k \geq 1$ and $\frac{s}{k} \leq \frac{1}{2}$. It is true that
	\begin{equation} \label{eq_appendix_1}
	1 = \left( \frac{s}{k} + \left( 1 - \frac{s}{k} \right) \right)^{k} \geq \sum_{i=0}^{s} {{k}\choose{i}} \left(\frac{s}{k}\right)^i \left(1-\frac{s}{k}\right)^{k-i} .
	\end{equation}
	Define $ A_i=\left(\frac{s}{k}\right)^i \left(1-\frac{s}{k}\right)^{k-i} $ for $i \in [0,s]$. Then
	\begin{multline} \label{eq_appendix_2}
	\log_2 A_i = i \log_2\left(\frac{s}{nk}\right) + \left(k-i\right) \log_2\left(1-\frac{s}{k}\right) \geq
	\\
	s \log_2\left(\frac{s}{k}\right) + \left(k-s\right) \log_2\left(1-\frac{s}{k}\right) = -k h\left(\frac{s}{k}\right).
	\end{multline} 
	Using (\ref{eq_appendix_1}) and (\ref{eq_appendix_2}) yields, after some algebra, $ 2^{k h\left(\frac{s}{k}\right)} \geq \sum_{i=0}^{s} {{k}\choose{i}} $.

\bibliographystyle{IEEEtran}
\bibliography{references}

\balance

\end{document}